\documentclass[useAMS,usegraphicx]{mn2e}
\newcommand{\etal}{et~al.\ }

\newcommand{\CIV}{C~{\sc iv}}

\newcommand{\FeII}{Fe~{\sc ii}}
\newcommand{\FeIII}{Fe~{\sc iii}}

\newcommand{\MgII}{Mg~{\sc ii}}

\newcommand{\NV}{N~{\sc v}}

\newcommand{\kms}{\hbox{km~s$^{-1}$}}

\newcommand{\lumind}{\hbox{ergs~s$^{-1}$~Hz$^{-1}$}}

\begin{document}

\def\sarc{$^{\prime\prime}\!\!.$}
\def\arcsec{$^{\prime\prime}$}
\def\arcmin{$^{\prime}$}
\def\degr{$^{\circ}$}
\def\seco{$^{\rm s}\!\!.$}
\def\ls{\lower 2pt \hbox{$\;\scriptscriptstyle \buildrel<\over\sim\;$}}
\def\gs{\lower 2pt \hbox{$\;\scriptscriptstyle \buildrel>\over\sim\;$}}

\title[Fractions of Low-Ionization BALQSOs]{The Intrinsic Fractions and Radio Properties of Low Ionization
Broad Absorption Line Quasars}
\author[X.\ Dai, F.\ Shankar, \& G.\ R.\ Sivakoff]{Xinyu Dai$^{1}$\thanks{E-mail:$\;$dai@nhn.ou.edu}, Francesco Shankar$^{2}$, and Gregory R. Sivakoff$^{3}$\\
$1$Homer L.\ Dodge Department of Physics and Astronomy, University of Oklahoma, Norman, OK 73019, USA\\
$2$Max-Planck-Institut f$\ddot{u}$r Astrophysik, Karl-Schwarzschild-Str. 1, D-85748 Garching, Germany\\
$3$Department of Astronomy, University of
Virginia, Charlottesville, VA 22904, USA}

\date{}
\pagerange{\pageref{firstpage}--
\pageref{lastpage}} \pubyear{2010}
\maketitle
\label{firstpage}

\begin{abstract}
    Low-ionization (\MgII, \FeII, \FeIII) broad absorption line quasars (LoBALs) probe a relatively obscured quasar population, and could be at an early evolutionary stage for quasars.  We study the intrinsic fractions of LoBALs using the SDSS, 2MASS, and FIRST surveys.  We find that the LoBAL fractions of the near infra-red (NIR) and radio samples are approximately 5--7 times higher than those measured in the optical sample.  This suggests that the
fractions measured in the NIR and radio bands are closer to the intrinsic fractions of the populations, and that the optical fractions are significantly biased due to obscuration effects, similar to high-ionization broad absorption line quasars (HiBALs).  We also find that the LoBAL fractions
decrease with increasing radio luminosities, again, similar to HiBALs.
    In addition, we find tentative evidence for high fractions of LoBALs at high NIR luminosities, especially for FeLoBALs with a fraction of $\sim$18 per cent at $M_{K_s} < -31$~mag.  This population of NIR luminous LoBALs may be at an early evolutionary stage of quasar evolution.
    We use a two-component model of LoBALs including a pure geometric component and a luminosity dependent component at high NIR luminosities, and obtain better fits than those from a pure geometric model.
    Therefore, the LoBAL population can be modelled as a hybrid of both the geometric and evolutionary models, where the geometric component constitutes 3.4$\pm$0.3, 5.8$\pm$0.4, and 1.5$\pm$0.3 per cent of the quasar population for BI-LoBALs, AI-LoBALs, and FeLoBALs, respectively.
    Considering a population of obscured quasars that do not enter the SDSS survey, which could have a much higher LoBAL fraction, we expect that intrinsic fraction of LoBALs could be even higher.
\end{abstract}


\section{Introduction}
Broad absorption line quasars (BALQSOs) are a sub-sample of quasars
exhibiting blue-shifted absorption troughs (e.g., Weymann et al.\ 1991).
In the traditional definition of Weymann et al.\ (1991), absorption
troughs must be at least 2000~\kms\ wide excluding the first 3000~\kms\ region blue-ward from the emission lines to classify quasars as BALQSOs.
Less strict definitions have also been used, for example with a requirement of a trough to be at least 1000~\kms\ wide (e.g., Trump et al.\ 2006).
BALQSOs can also be further divided into a population containing absorption troughs from only high-ionization state species (e.g., \CIV\ and \NV; HiBALs)
and a population that exhibits absorption troughs in low-ionization species (e.g., \MgII\ and \FeII; LoBALs).
The majority of BALQSOs are HiBALs.
In fact, all LoBALs also contain the high-ionization troughs in their spectra (e.g., Weymann et al.\ 1991; Trump et al.\ 2006).
Besides the presence of low-ionization troughs, the optical continua of LoBALs are more reddened compared to HiBALs, suggesting stronger dust extinction (e.g, Sprayberry \& Foltz 1992; Reichard et al.\ 2003).
In X-rays, LoBALs also have higher gas absorption column densities than HiBALs (e.g., Green et al.\ 2001; Gallagher et al.\ 2002).
Therefore, LoBALs probe a relatively obscured quasar population.
The origin of a small LoBAL fraction in quasars is unclear, and it has been attributed to geometric effects (e.g., Elvis et al.\ 2000) or evolutionary effects (e.g., Voit et al.\ 1993), like the BALQSO population in general.
There are several tentative arguments supporting the view that LoBALs are young quasars at a stage of blowing out obscuring materials.
First, several early studies of LoBAL fractions in the infra-red band showed larger fractions (e.g., Boroson \& Mayers 1992) and associations with ultra-luminous infra-red galaxies (e.g., L{\'{i}}pari et al.\ 1994; Canalizo \& Stockton 2000).
Second, a few optical spectral analyses suggested that the covering fraction of the LoBAL wind is large (e.g., Voit et al.\ 1993; Casebeer et al.\ 2008).
Third, some radio spectra of LoBALs resemble those of compact steep spectrum or gigahertz peaked spectrum sources, which are also candidates for young quasars (e.g., Montenegro-Montes et al.\ 2008; Liu et al.\ 2008).
In particular, the LoBALs containing Fe absorption troughs (FeLoBAL) are viewed as the most promising candidates for young quasars (e.g., L{\'{i}}pari et al.\ 2009).
The fractions of LoBALs are important constraints on the origin of the LoBAL populations.

Before studying the intrinsic fractions of LoBALs, it is important to compare the measurements of the intrinsic fractions of BALQSOs in the quasar population.
Recently, a series of studies emerged on this topic. 
Dai, Shankar, and Sivakoff (2008a) studied the SDSS BALQSO (Trump et al.\ 2006) fractions in the 2MASS bands (Skrutskie et al.\ 2006), finding that the BALQSO fractions in the near-infrared (NIR) are twice those found in the optical band. In particular, we found the BALQSO fraction to be 20$\pm$2 per cent for the traditional BALQSOs that satisfy the stricter Weymann et al. (1991) definition
and 43$\pm$2 per cent for the relaxed definition of Trump et al. (2006) that requires less broad absorption troughs.
Dai et al.\ (2008a) argued that the BALQSO fractions measured in the NIR bands are closer to the intrinsic fractions, 
 based on the observations that significant obscuration is associated with BALQSOs in the optical bands (e.g., Reichard et al.\ 2003), confirming the earlier estimate of Hewett \& Foltz (2003).
This result was confirmed by several studies, such as Ganguly \& Brotherton (2008) using a different SDSS BALQSO catalogue, Maddox et al.\ (2008) using the deeper UKIDSS survey, Shankar et al. (2008a) in the radio band, and Knigge et al.\ (2008) by correcting the fraction in the optical bands directly.
In particular, Ganguly \& Brotherton (2008) extended the study to include narrow and associated absorbers and found the overall outflowing AGNs to be 60 per cent of the total quasar population.
Recently, Allen et al.\ (2011) claimed an even larger intrinsic fraction for the traditional BALQSOs of 41 per cent, by including the additional fraction of missing quasars that do not enter the SDSS survey.
In particular, Allen et al.\ (2011) found that the completeness for BALQSOs and non-BALQSOs in SDSS is very similar at $z < 2.1$ and $z > 3.6$, but can differ at other redshifts, e.g., for $z\sim2.6$ and $z\sim3.5$.

The larger fraction of BALQSOs makes the AGN wind a more promising candidate responsible for the feedback energy that is
needed to explain the co-evolution between black holes and their host galaxies.
Understanding evolutionary versus geometric models of AGNs
can not only probe the AGN feedback, but also constrain the nature
of the feedback, whether it is kinetic from winds (e.g., Granato et al.\ 2004, Shankar et al.\ 2006, 2008b)
or thermal (e.g., Di Matteo et al.\ 2005; Hopkins et al.\ 2006).
If BALQSOs provide the majority of the feedback energy, the feedback mechanism will be kinetic.
Motivated by the results from BALQSOs, and the larger obscuration of LoBALs compared to HiBALs, we expect that the optical fractions for LoBALs are also biased low.
This effect has already been noticed when only a few LoBALs were observed (Sprayberry \& Foltz 1992); however, their study was limited by their small sample size.
The large sample size enabled by SDSS warrants a new study on the intrinsic fractions of LoBALs.

The radio properties of BALQSOs provide additional constraints on the nature of these objects.
In an early study, Stocke et al.\ (1992) found no radio-loud BALQSOs within 68 BALQSOs.
Later studies showed that radio emission is present in BALQSOs (Francis et al.\ 1993; Brotherton et al.\ 1998; Becker et al.\ 2000); however, these BALQSOs are mostly radio-moderate.
Matching the SDSS BALQSO catalogue in the FIRST survey (Becker et al.\ 1995), Shankar et al.\ (2008a) quantified the dependence of the BALQSO fraction on radio luminosities.
We found that the BALQSO fraction drops at high radio luminosities confirming earlier claims of such an effect (Stocke et al.\ 1992; Becker et al.\ 2001).
In addition, Shankar et al.\ (2008a) also found that the BALQSO fraction at the low radio luminosity range is consistent with the NIR fraction of BALQSOs of Dai et al.\ (2008a).
This result further supports the view that the NIR BALQSO fraction is close to the intrinsic fraction (modulo corrections to the parent SDSS quasar selection), since there is also little absorption in the radio bands.
The drop of the BALQSO fractions at high radio luminosities can be naturally explained under a geometric model of BALQSOs.
If the radio emission has a preferred orientation, which is usually considered in the polar direction, the drop indicates that BALQSOs are less frequent in these viewing angles.
Using a unification model between radio-loud and radio-quiet quasars (e.g., Urry \& Padovani 1995), we were able to successfully reproduce the trend, thus explaining the majority of the BALQSOs with radio emission under a geometric model.
Exceptions still exist, such as the polar radio-loud BALQSOs, which were identified based on the radio variability that implies too large brightness temperatures unless the radio emission is relativistic (e.g., Zhou et al.\ 2006; Ghosh \& Punsly 2007).
Cold polar outflows are also present in blazars (e.g., Dai et al.\ 2008b), and they could be related to polar BALQSOs as the outflow continuously extends close to the polar axis.
However, these objects are rare, and their implication for the total BALQSO population is still uncertain.

In this paper,
we study the intrinsic fraction of LoBALs by correlating SDSS quasars with detections in the NIR and radio bands.
We also explore their radio properties and compare with BALQSOs to test whether LoBALs can be explained under a geometric or evolutionary model.
We assume that $H_0 = 70~\rm{km~s^{-1}~Mpc^{-1}}$, $\Omega_{\rm m} = 0.3$,
and $\Omega_{\Lambda}= 0.7$ throughout the paper.

\begin{figure*}
    \includegraphics[width=17.5truecm]{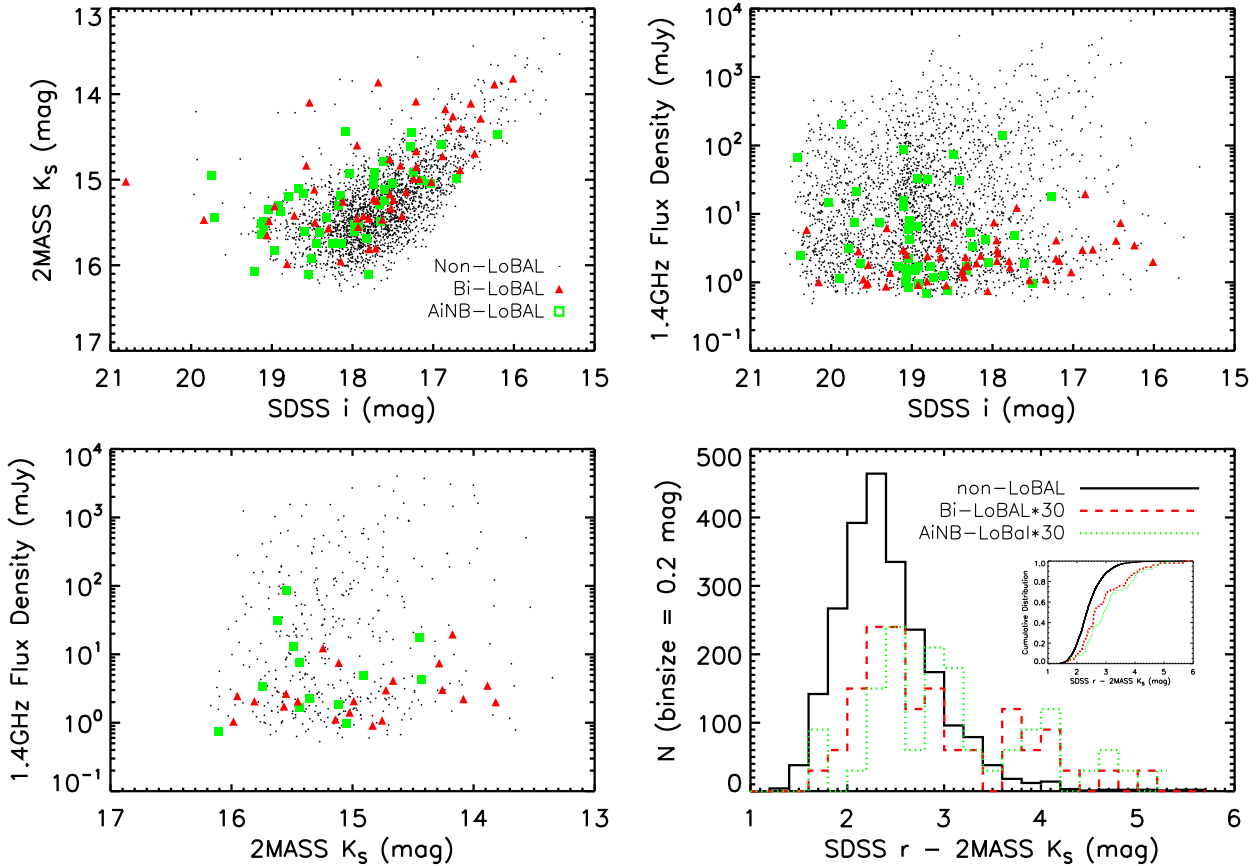}
    \caption{(top left) 2MASS $K_s$ magnitude versus SDSS $i$ magnitude for SDSS DR3 QSOs that are detected in all of the
    $J$, $H$, and $K_s$ bands in the redshift range of $0.5 \le z \le 2.15$. The subsamples of QSOs that are not LoBALs
    (non-LoBALs; black dots), QSOs that satisfied the traditional Weymann et al.\ definition (BI-LoBALs; red triangles), and
    QSOs that satisfied the relaxed BAL definition of Trump et al.\ (2006), but did not satisfy the traditional definition
    (AINB-LoBALs; green squares) are displayed separately. (top right) SDSS $i$ magnitude versus 1.4~GHz flux density.
    (bottom left) 2MASS $K_s$ magnitude versus 1.4~GHz flux density.  (bottom right) SDSS $r -$ 2MASS $K_s$ colour for the
     three samples, where the histograms for the two LoBAL samples are multiplied arbitrarily by 30 for clarity.  The LoBALs are
     significantly redder than the non-LoBAL population.  The inset in the bottom right panel shows the cumulative distributions
     of the colour for the three samples.  The K--S test results indicate that both of the LoBAL samples differ from the non-LoBAL
     sample with significances greater than 99.996 per cent, and that the two LoBAL samples are not significantly different from each
     other. \label{fig:one}}
\end{figure*}

\begin{figure*}
    \includegraphics[width=17.5truecm]{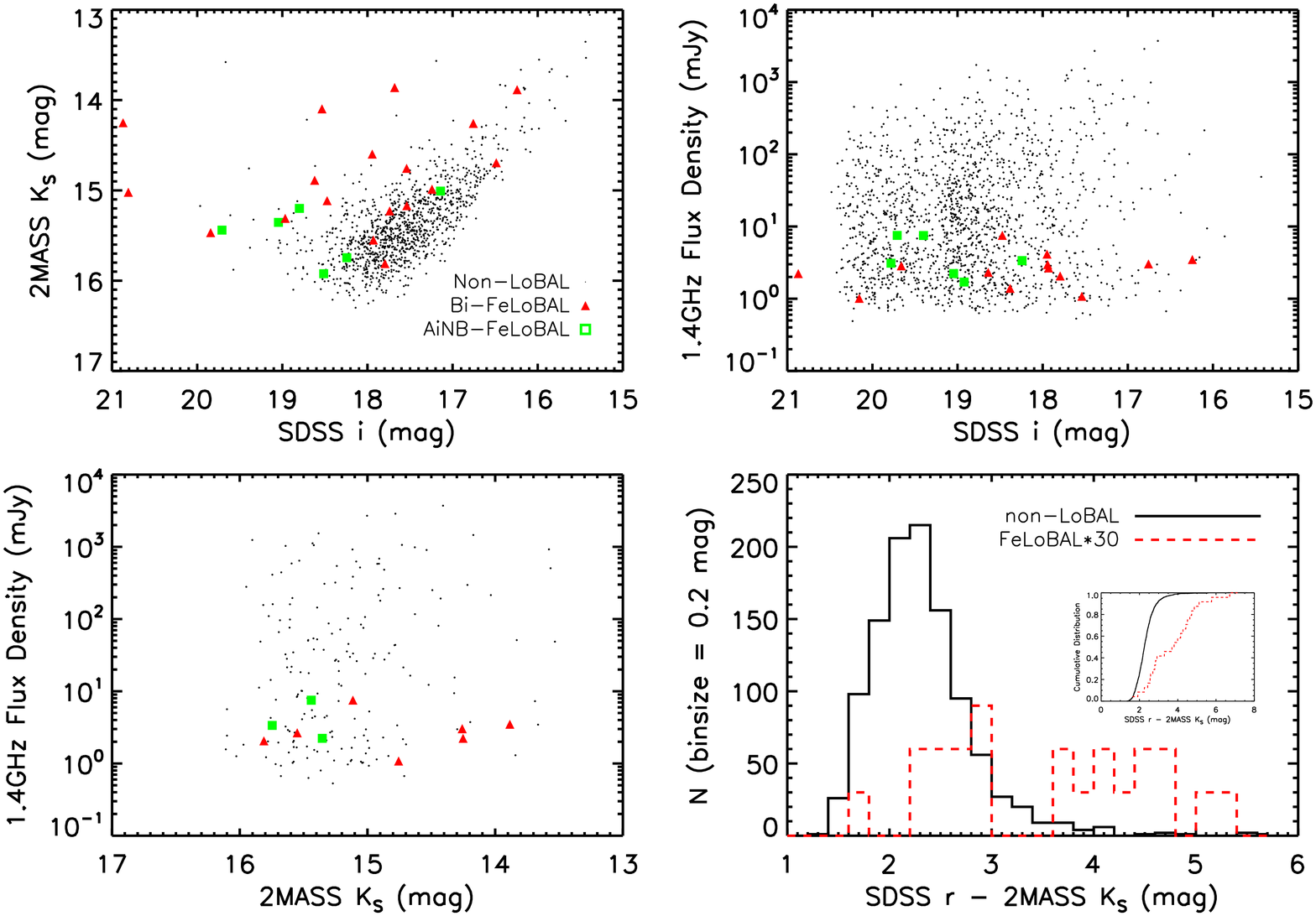}
    \caption{Same plots as Fig.~\ref{fig:one}, but for FeLoBals in the redshift range of $1.19 \le z \le 2.24$.  The FeLoBALs are even more significantly redder than the non-LoBAL population.  The K--S test result indicates that the $r-K_s$ colours of the FeLoBAL population differ from those of the non-LoBAL population with a significance greater than $1-10^{-8}$. \label{fig:two}}
\end{figure*}

\section{Sample Selection}
We select quasars from the SDSS DR3 quasar catalogue (Schneider et al. 2005) and the
SDSS-BALQSO catalogue (Trump et al.\ 2006, T06 hereafter), where the authors also separately tagged the LoBAL population.  In particular, we study quasars in the redshift
range of $0.5 \le z \le 2.15$, where the \MgII\ absorption line falls
in the observed-frame optical band pass.
Note that, in this redshift range, the completeness of BALQSOs and non-BALQSOs is very similar for the SDSS survey (Allen et al. 2011).
For the FeLoBAL sample, we study the quasars in the redshift range of $1.19 \le z \le 2.24$, which is the redshift range of FeLoBALs identified in T06.
Following Dai et al.\ (2008a) and Shankar et al.\ (2008a),
we match the quasars using a
2\arcsec\ radius to entries from the full 2MASS release, which extends to fluxes below the official completeness level of the 2MASS All-Sky Point Source Catalogue.
To ensure that the matched database entries represent detections, we
require an in-band detection (rd\_flg != 0) that there is no confusion,
contamination (cc\_flg == 0), or blending (bl\_flg $\le$ 1) in the source, and that no source is near an extended galaxy (gal\_contam == 0 and ext\_key is
null).
The 99 per cent completeness levels of the database are $J=16.1$, $H=15.5$, and
$K_{s}=15.1$~mag.
Recently, several new BALQSO catalogues have been published using more recent SDSS data releases (Gibson et al.\ 2009; Scaringi et al.\ 2009; Allen et al.\ 2011). In particular, the catalogue of Gibson et al.\ (2009, G09 hereafter) also provides identifications of traditional \MgII\ LoBALs in the redshift range of $0.55 \le z  \le 2.15$.
We also briefly present the analysis to this sample as a comparison.  
We mainly present results based on the T06 sample, unless mentioned otherwise. 

Following Shankar et al.\ (2008a, 2010),
 we build a full FIRST-SDSS cross-correlation
catalogue, containing all the detected radio components within
30\arcsec of an optical quasar. As our reference, following Schneider
et al. (2005), we primarily present results for the FIRST-SDSS
catalogue with radio counterparts identified within 2\arcsec.
We also test our results by enlarging the radio
matches to 5\arcsec, and the results are consistent with those from the 2\arcsec\ matching. In addition, many of the optical sources in
SDSS are associated with more than one radio component in FIRST within
30\arcsec, as expected if these sources are extended with jets
and/or lobes separated from the central source.
We use the sum of the flux
densities as a proxy for the total radio luminosity of these sources.
We require that at least one component is within our matching radius for the cases of multiple source matches.

In the rest of the paper, we refer to LoBALs that satisfy the traditional definition of Weymann et al.\ (1991) as
``BI-LoBALs'', to LoBALs that are selected from T06 using the relaxed AI definition as
``AI-LoBALs'', to LoBALs that satisfy the relaxed AI definition but not the traditional definition as
``AINB-LoBALs'', and to the rest of the sample without LoBAL features as ``non-LoBALs''.
Some studies have found that the AINB sample has properties besides the absorption troughs which differ from
the BI sample, which may indicate a separate population (e.g., Knigge et al,\ 2008, Shankar et al.\ 2008a).
We note that at the redshift ranges we study, the SDSS spectra cannot separate all the HiBALs from non-BALs because the high-ionization troughs are not redshifted into the optical spectral range.
We also refer to the ``2MASS sample'' as quasars detected in all of the $J$, $H$, and $K_s$ bands, and to the ``$K_s$ complete sample'' as quasars with $K_s < 15.1$~mag.
We test our results using the two samples, and they are not qualitatively different.  Because of the scarcity of LoBALs, we choose to mainly present the results from the 2MASS sample to reduce the uncertainties of our fraction measurements.

\section{Results}

\subsection{Distributions of LoBALs in the Optical, NIR, and Radio Bands}
In three panels of Fig.~\ref{fig:one}, we plot the SDSS~$i$ magnitude, 2MASS $K_s$ magnitude, and FIRST 1.4~GHz flux density against each other for three samples, BI-LoBALs, AINB-LoBALs, and non-LoBALs.
The distributions of LoBAL samples are different from the non-LoBAL sample in all three cases.
In the $K_s$ versus $i$ magnitude plot (Fig.~\ref{fig:one}, top left), the quasars are concentrated in a linear relation between the $i$ and $K_s$
magnitudes with scatter.
We can see that the LoBALs in the plot are systematically redder than the non-LoBAL population, as expected since the optical spectra of LoBALs show extra dust extinction compared to HiBALs and non-BALs (e.g., Reichard et al.\ 2003).
We note that some of the scatter is due to quasar variability, since the SDSS and 2MASS data are taken from different epochs.
However, this should not affect the mean colour difference between the LoBALs and non-LoBALs in the plot.
In the two panels involving radio flux in Fig.~\ref{fig:one}, we clearly see that the LoBALs are less populated in the high radio flux regime.
In the bottom-right panel of Fig.~\ref{fig:one}, we show the histograms of the SDSS~$r -$2MASS~$K_s$ colour for the BI-LoBALs, AINB-LoBALs, and non-LoBALs in our sample as another example to demonstrate the redder colour of the LoBAL population.
We also show the cumulative distribution of the $r-K_s$ colour in the inset, and perform the Kolmogorov-Smirnov (K--S) test on the distributions.
The K--S test results show that the AI-LoBALs of our sample are significantly different from the non-LoBALs with a very small probability ($3\times10^{-11}$) for the two populations to arise from the same parent distribution.
The situation is the same when we compare the sub-populations of the AI-LoBALs (BI-LoBALs and AINB-LoBALs) with the non-LoBALs, where we obtain K--S probabilities of $4\times10^{-5}$ and $4\times10^{-8}$.
Comparing the BI-LoBALs and AINB-LoBALs in the plot, we find that the two colour distributions are not significantly different with a K--S probability of 0.28.
In Fig.~\ref{fig:two}, we compare the properties of the FeLoBALs and non-LoBALs  of our sample using the same axes as in Fig.~\ref{fig:one}.
We conclude that FeLoBALs are even more reddened than LoBALs, as expected.
The radio powers of the FeLoBALs in the plot are similar to those of the BI-LoBALs.
Ideally, these tests should be performed in narrower redshift bins, since there is quasar colour evolution with redshift; however, we are limited by the small sample size, and defer this analysis to future studies.

\subsection{The Intrinsic Fraction of LoBALs\label{sec:intr}}

\begin{figure*}
    \includegraphics[width=8.5truecm]{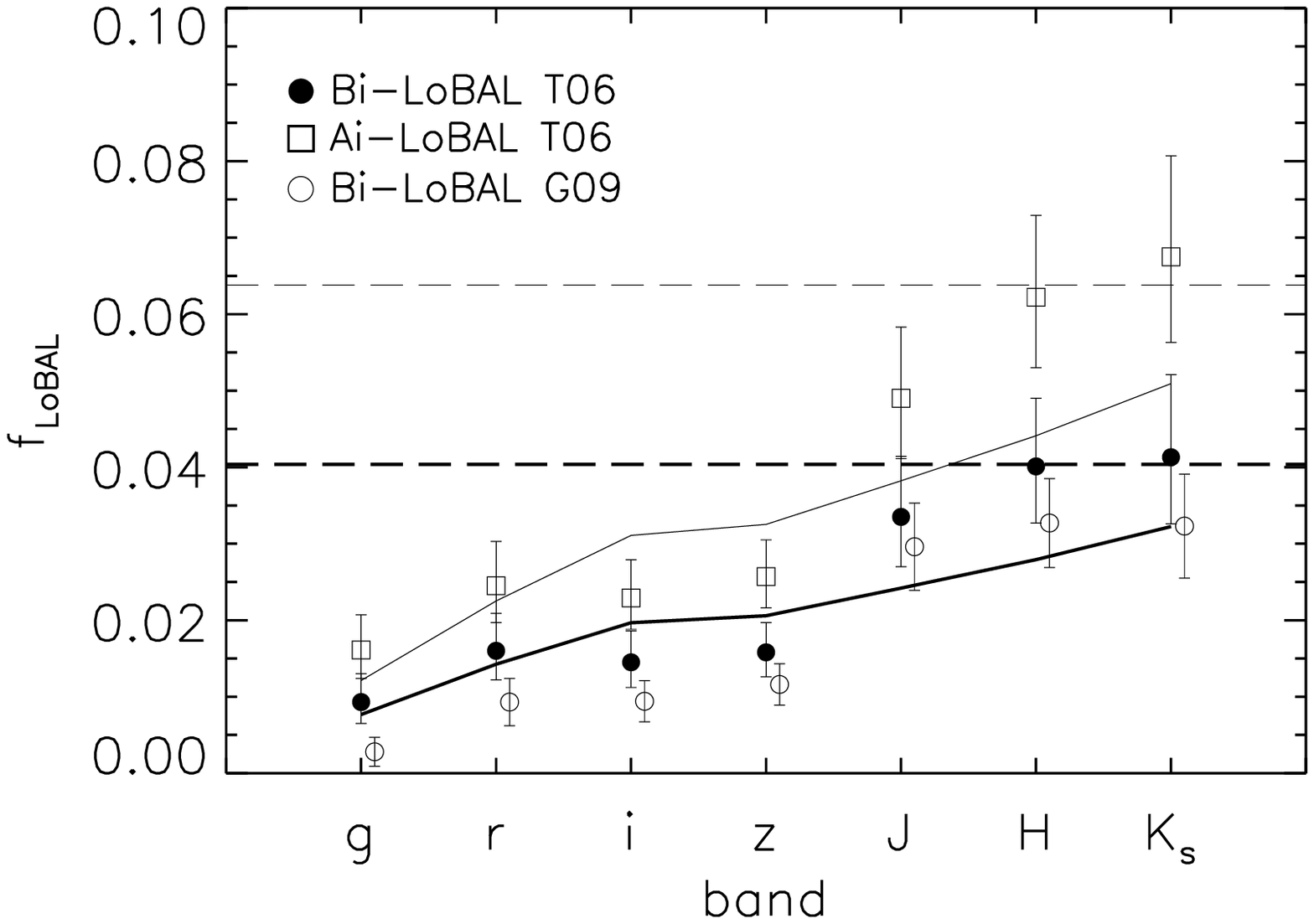}
    \includegraphics[width=8.5truecm]{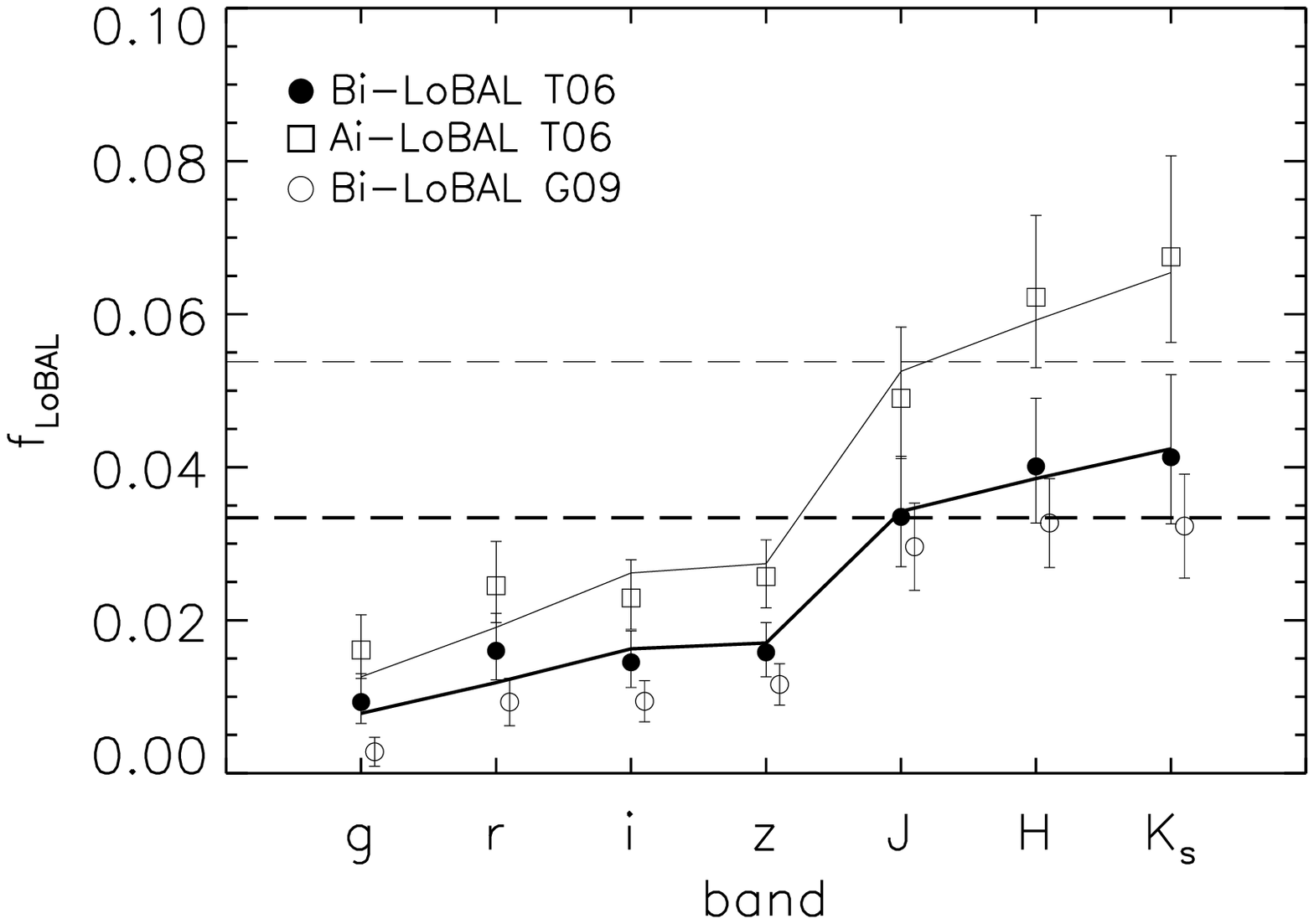}
    \caption{(left) Observed LoBAL fractions in optical and infrared bands.  The filled circles and open squares are the observed BI-LoBAL and AI-LoBAL fractions from T06, and the open circles are from G09, where we shift the G09 data slight to the right for clarity.   The BAL fractions increase from the blue to red bands, and the fractions can be modelled assuming there is significant obscuration in LoBALs. The heavy dashed line is the model intrinsic fraction for BI-LoBALs of T06, and the light dashed line is for that of AI-LoBALs of T06. The solid lines are the modeling results for the observed fractions. (right) We add an additional component of LoBALs at high luminosities to our model, where we obtain better fits to the data.  The dashed lines show the model intrinsic fraction for the geometric component of LoBALs. \label{fig:thr}}
\end{figure*}

\begin{figure*}
    \includegraphics[width=8.5truecm]{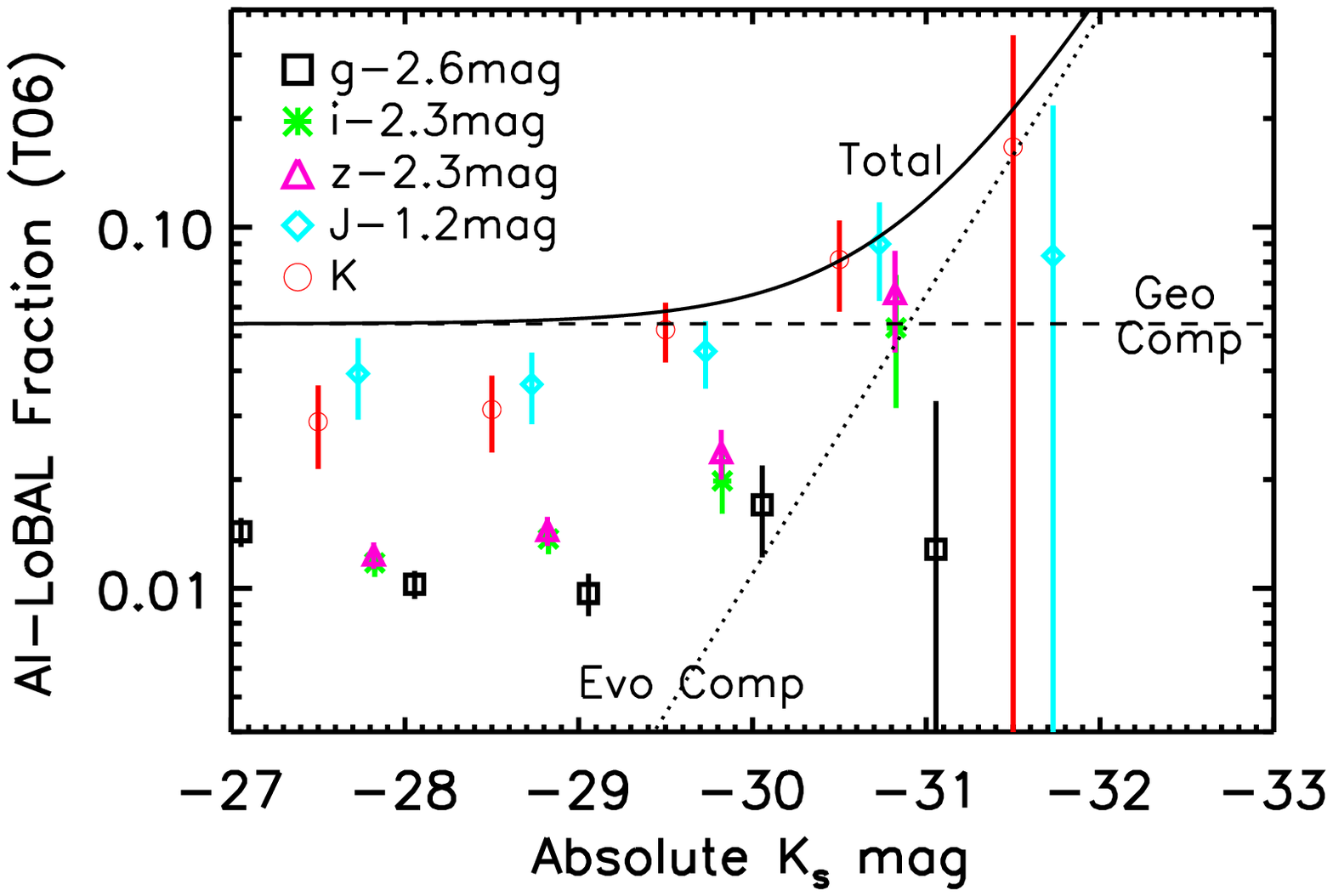}
    \includegraphics[width=8.5truecm]{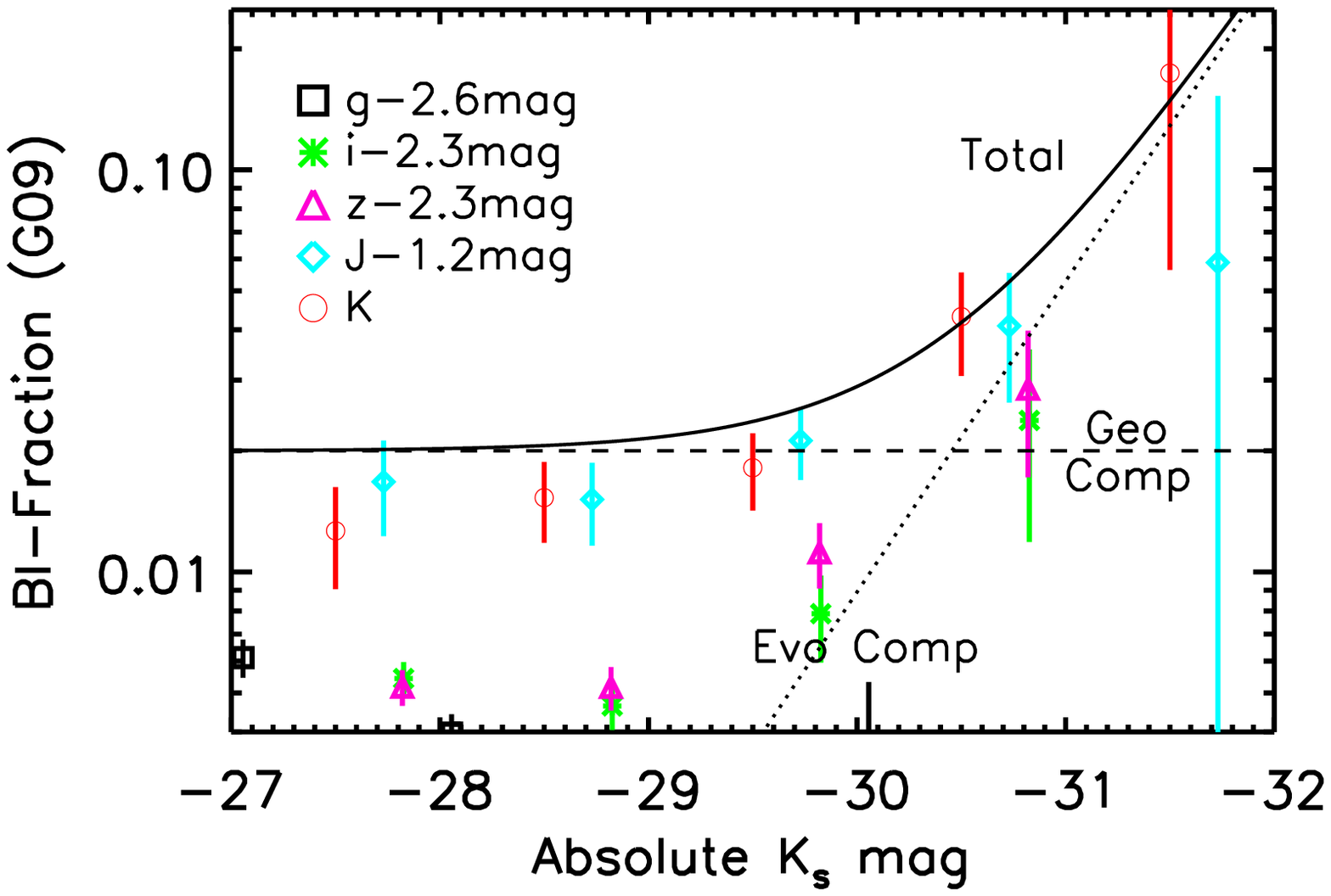}
    \caption{ (left) The observed AI-LoBAL fractions as a function of optical and infrared luminosities.  The $g, i, z$, and $J$ band data are shifted right horizontally according to the mean colour differences with respect to the $K_s$ band. The fractions increase from the blue to red bands because of the selection biases, especially in shorter wavelength bands.  We also find that the AI-LoBAL fractions increase from low to high luminosities, especially in the redder bands. 
    The solid line shows the model intrinsic fractions from Model II, where the dashed and dotted lines are the geometric and evolutionary component, respectively.  (right) Same plot but for BI-LoBAL fractions from the G09 sample. \label{fig:fou}}
\end{figure*}

\begin{figure*}
    \includegraphics[width=8.5truecm]{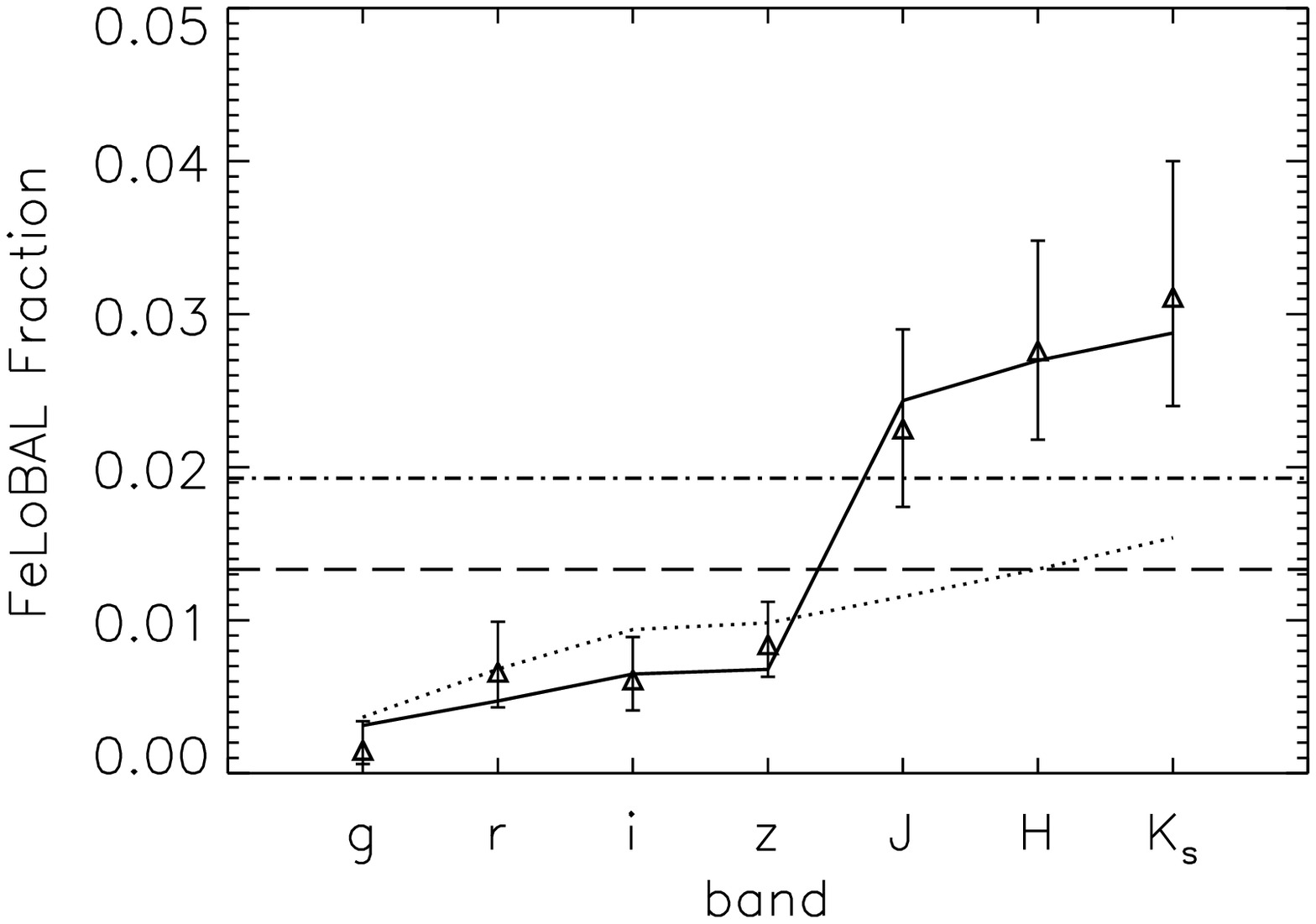}
    \includegraphics[width=8.5truecm]{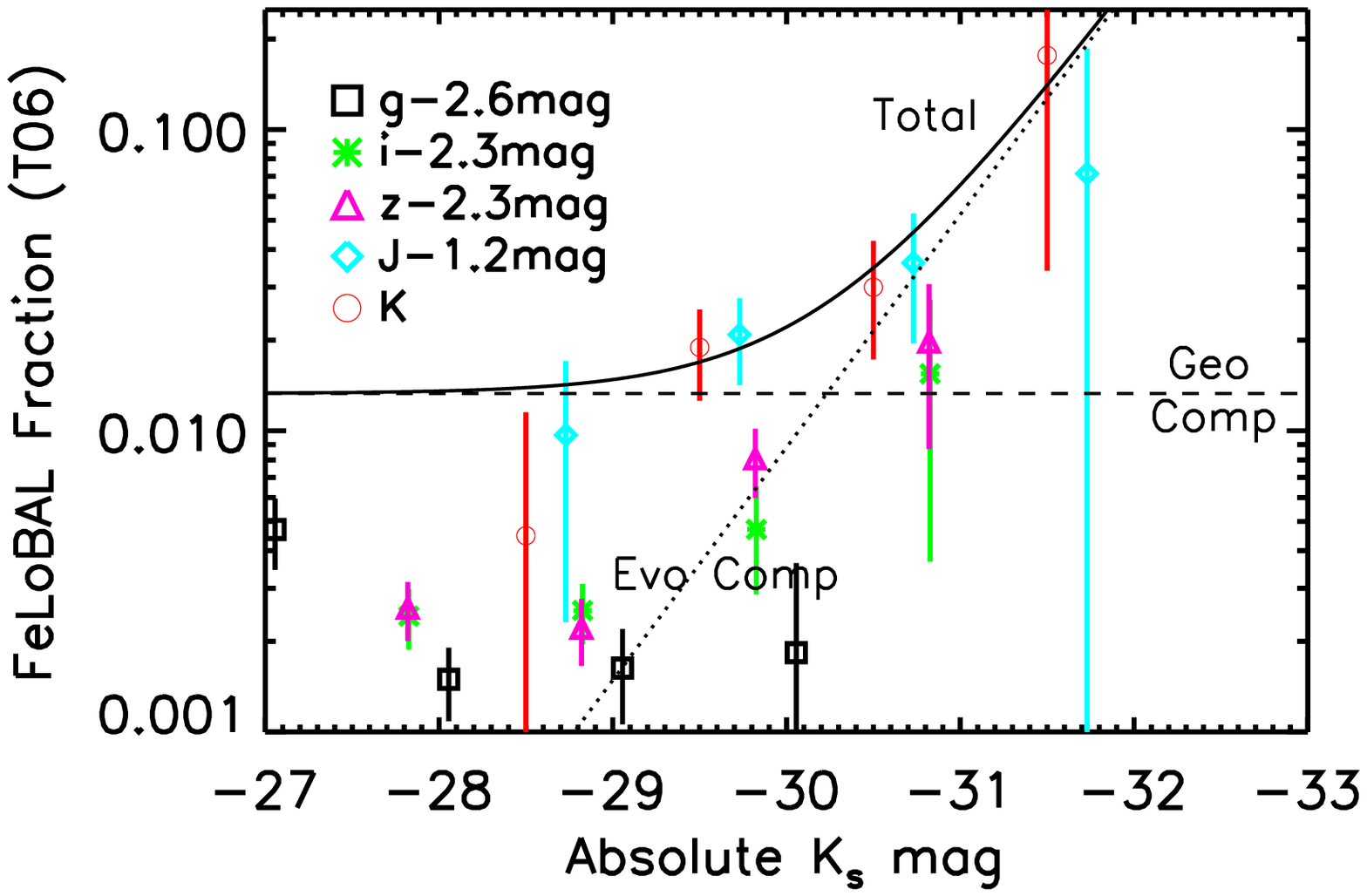}
    \caption{(left) Observed FeLoBAL fractions from the $g$ to $K_s$ bands.  The triangles are the data.  The dotted line is the best fit from a pure geometric model, and dot-dashed line shows the intrinsic fraction from this model.  The solid line is the best fit from Model II (geometric plus evolutionary), and the dashed line shows intrinsic fraction for the geometric component.  (right) Observed FeLoBAL fractions as a function of optical and infrared luminosities.  The solid line is the model intrinsic FeLoBAL fraction from Model II, and the dashed and dotted lines are the geometric and evolutionary component, respectively.  \label{fig:fiv}}
\end{figure*}

\begin{table*}
\caption{Fractions of LoBALs in the 2MASS and SDSS Bands\label{tab:num}}
\begin{center}
\begin{tabular}{lccccccccccc}
\hline
\hline
{} & Sample & {Redshift Range} & {$u$} & {$g$} & {$r$} & {$i$} & {$z$} & {$J$} & {$H$} & {$K_s$} \\
\hline
Limiting Mag  &  &  & $-$26.7 & $-$26.9 & $-$27.1 & $-$27.2 & $-$27.2 & $-$28.3 & $-$28.9 & $-$29.5 \\
\hline
N (Total) &      T06 & 0.50--2.15     & 1460 & 1181 & 1062 & 1312 & 1518 & 776 & 724 & 533 \\
N (BI-LoBAL) &      &   &   5  &   11 &   17 &  19  &   24 &  26 &  29 &  22 \\
f (BI-LoBAL) &    &     & $0.3^{+0.2}_{-0.2}$ & $0.9^{+0.4}_{-0.3}$ & $1.6^{+0.5}_{-0.4}$     & $1.5^{+0.4}_{-0.3}$     & $1.6^{+0.4}_{-0.3}$ & $3.4^{+0.8}_{-0.7}$    &$4.0^{+0.9}_{-0.7}$   &  $4.1^{+1.1}_{-0.9}$   \\
N (AI-LoBAL) &      &   &  12  &   19 &   26 &  30  &   39 &  38 &  45 &  36 \\
f (AI-LoBAL) &    &     & $0.8^{+0.3}_{-0.2}$     &  $1.6^{+0.5}_{-0.4}$    &  $2.5^{+0.6}_{-0.5}$    &  $2.3^{+0.5}_{-0.4}$    &  $2.6^{+0.5}_{-0.4}$    &  $4.9^{+0.9}_{-0.8}$   &  $6.2^{+1.1}_{-0.9}$   &  $6.7^{+1.3}_{-1.1}$   \\
\hline
N (Total)  &  & 1.19--2.24 & 1584 & 1366 & 1207 & 1486 & 1783 & 843 & 798 & 610 \\
N (FeLoBAL)  &  &       &    1 &    2 &    8 &    9 &   15 &  19 &  22 &  19 \\
f (FeLoBAL)  &    &     &  $0.06^{+0.02}_{-0.01}$    &  $0.2^{+0.2}_{-0.1}$    &  $0.7^{+0.3}_{-0.2}$    &  $0.6^{+0.3}_{-0.2}$    &   $0.8^{+0.3}_{-0.2}$   &  $2.3^{+0.7}_{-0.5}$   &  $2.8^{+0.7}_{-0.6}$   &  $3.1^{+0.9}_{-0.7}$   \\
\hline
N (Total)  & G09  & 0.55--2.15 & 2225 & 1817 & 1621 & 2030 & 2337 & 1252&1313 & 991 \\
N (BI-LoBAL) &  &          &   2  &   5  &  15  &  19  &  27  & 37  & 43  & 32  \\
f (BI-LoBAL) & &        &  $0.09^{+0.12}_{-0.06}$    &  $0.3^{+0.2}_{-0.1}$    &  $0.9^{+0.3}_{-0.2}$    &  $0.9^{+0.3}_{-0.2}$    &  $1.2^{+0.3}_{-0.2}$    &   $3.0^{+0.6}_{-0.5}$  &   $3.3^{+0.6}_{-0.5}$  &   $3.2^{+0.7}_{-0.6}$  \\
\hline
\end{tabular}
\end{center}
\raggedright Notes: The fractions are in per cents.
\end{table*}

\begin{table*}
\caption{Modeling Results for the Intrinsic Fractions of LoBALs Using the 2MASS and SDSS Data\label{tab:ifrac}}
\begin{center}
\begin{tabular}{lccccc}
\hline
\hline

{} & Model & AI T06 & BI T06 & BI G09 & Fe T06 \\
\hline
Intrinsic Fraction (per cent) & I & 6.4$\pm$0.8 & 4.0$\pm$0.6 & 2.8$\pm$0.4 & 1.9$\pm$0.4 \\
$\chi^2/dof$ &  & 13.4/6 & 9.0/6 & 21.6/6 & 16.9/6 \\
Intrinsic Geometric Fraction (per cent) & II & 5.4$\pm$0.5 & 3.3$\pm$0.4 & 2.0$\pm$0.3 & 1.3$\pm$0.3 \\
$\alpha$ & & 0 & 0 & 0 & 0  \\
$\Phi_{0, EVO} / \Phi_{0, Total}$ & & 6.1$\times10^{-5}$ & 4.2$\times10^{-5}$ & 5.0$\times10^{-5}$ & 4.9$\times10^{-5}$ \\
$\chi^2/dof$ &  & 2.7/4 & 1.5/4 & 2.6/4 & 2.4/4 \\
\hline
\end{tabular}
\end{center}
\raggedright Notes: Model I is a pure geometric model, where the only free parameter is the intrinsic fractions of LoBALs.  Model II is a two components model, where we add an additional component (evolutionary) to Model I.  The parameter $\alpha$ is the power-law slope of the luminosity function of the evolutionary component for LoBALs, $\Phi_{EVO} \propto \Phi_{0, EVO} L^{\alpha}$.  The parameter $\Phi_{0, EVO} / \Phi_{0, Total}$ is the normalization ratio between the luminosity function of the evolutionary component for LoBALs and total quasars.  Both $\alpha$ and $\Phi_{0, EVO} / \Phi_{0, Total}$ are unconstrained from the data.  We list one set of possible solutions that produce good fits in this table.
\end{table*}

We present the fraction of LoBALs (BI-LoBALs, AI-LoBALs, and FeLoBALs) measured in the SDSS and 2MASS $g, r, i, z, J, H$, and $K_s$ bands in Figs.~\ref{fig:thr} and \ref{fig:fiv} (left).
We also list the numbers and fractions of LoBALs in Table~\ref{tab:num}, where we use Gehrels's statistics to estimate the uncertainties (Gehrels 1986).
We first calculate the absolute magnitudes of the quasars in the SDSS and 2MASS bands correcting for the Galactic extinction, but not correcting for the obscuration in the BALQSOs.
We then calculate the fraction in the $K_s$ band with an absolute magnitude limit of $M_{K_s} = -29.5$~mag.
For the rest of the bands, we set the limits so that the differences between the band limits are the same as their mean colour differences, i.e., $M_{X, lim} - M_{Y, lim} = \overline{M_X - M_Y}$.
We find that the LoBAL fractions are increasing from blue to red bands, similar to the results for BALQSOs (Dai et al.\ 2008a).
This trend can be naturally explained considering the significant obscuration in LoBALs compared to HiBALs and non-BALs (e.g., Sprayberry \& Foltz 1992), and the implication is that the 2MASS fractions are closer to the intrinsic fraction of these objects.
Comparing the BI-LoBAL samples from T06 and G09, we find that overall trend is consistent between the two samples.  However, there is a systematic offset between the two samples, where the T06 sample has higher fractions.

Following Dai et al.\ (2008a), we test the obscuration model by combining the quasar luminosity function (Richards et al.\ 2005) and the spectral differences between LoBALs, HiBALs, and non-BALs (Reichard et al.\ 2003),  Model I hereafter.
We assume that LoBALs, HiBALs, and non-BALs have the same intrinsic luminosity functions (Richards et al.\ 2005) with the only difference being in their normalizations.
Since LoBALs and HiBALs are obscured, their observed luminosity functions are horizontally shifted to be less luminous, where the shift corresponds to the mean obscuration caused by dust extinction and absorption.
We calculate the obscuration based on the SDSS composite spectral models (Reichard et al.\ 2003).
The shift is smaller in the red bands and larger in the blue bands.
Finally, we calculate the model observed fraction of LoBALs in the SDSS and 2MASS bands, and compare with the observations.
The only free parameter is the intrinsic fraction of LoBALs.
Please refer to Dai et al.\ (2008a) for the detailed model.
This assumes that the differential selection function for LoBALs and non-LoBALs in SDSS only depends on the different colours due to extinction and absorption troughs, which we believe is reasonable given our selected redshift range and the results of Allen et al.\ (2011).
The model results are presented in Fig.~\ref{fig:thr} (left), where the dashed lines are the intrinsic fractions of LoBALs and the solid lines connect the model results for the observed fraction of LoBALs.  The square and filled circle symbols are the observed AI-LoBAL and BI-LoBAL fractions from T06.
We also show the observed BI-LoBAL fractions from G09 as a comparison.
The model produces acceptable fits to the data with $\chi^2/dof = 1.5$ and 2.2 for BI-LoBALs and AI-LoBALs, and measures intrinsic fractions of LoBALs of $4.0\pm0.6$ per cent for BI-LoBALs and $6.4\pm0.8$ per cent for AI-LoBALs.
We note that the model fits the data from the bluer bands better than the redder bands, which is possible that there is a separate IR luminous LoBAL population contributing to the LoBAL fractions, which we will discuss in detail later.
The FeLoBAL fractions also show an increasing trend from the bluer $g$ band to the redder $K_s$ band (Fig.~\ref{fig:fiv} left).
We list the model results for using a pure geometric model in Table~\ref{tab:ifrac}.

We plot the AI-LoBAL, BI-LoBAL (G09) and FeLoBAL fractions as a function of NIR and optical luminosities in Figs.~\ref{fig:fou} and \ref{fig:fiv} (right).
In general, we find that the fractions are larger in the redder bands than bluer bands, consistent with the results found above.
We also find that there is an increasing trend in LoBAL fractions as a function of luminosity, especially in the redder 2MASS bands.
The bluer $g$ band does not show such a trend.
This trend, especially in the redder bands, is not observed in BALQSOs in general (Dai et al.\ 2008a).
Besides the bluer bands, which are more affected by the obscuration, the other bands have similar slopes for the increase of LoBAL fractions as a function of luminosity.
However, the data at the luminous end have large uncertainties.
Early IR surveys of LoBALs (with sample sizes $\sim$10) typically found LoBAL fractions of \gs10 per cent (e.g., Boroson \& Meyers 1992), higher than our intrinsic fraction.  It is possible that these early IR surveys have higher flux limits that can only probe the most IR luminous population, where we also find a large LoBAL fraction of \gs10 per cent.

It is possible that the increasing fraction with luminosity is a result of increasing signal-to-noise ratio (S/N) spectra for quasars, such that it is more likely to identify sources as BALQSOs.  This effect has been discussed previously (Knigge et al.\ 2008; G09; Allen et al.\ 2011).
However, our sample only probes the most luminous quasars after applying the 2MASS flux limits, and in this regime, the effect due to S/N is small.  For example, in Figure 11 of Allen et al.\ (2011), the \CIV\ BAL fraction is a constant for the most luminous quasars. 
Moreover, if the trend of increasing fraction with luminosity is really due to the S/N of quasar spectra, it indicates that the intrinsic LoBAL fractions are close to $\sim$18 percent.  This is not consistent with our Model I prediction after taking the obscuration effects into account.
In the rest of the paper, we assume that the S/N has a small effect on the observed fractions in the luminosity range of our sample. 

We add another LoBAL component to our geometric model (Model II), based on the larger observed LoBAL fractions at high IR luminosities and the large $\chi^2$ values from a pure geometric fit in Fig.~\ref{fig:thr} (left).  We call this component the evolutionary component.  We model the evolutionary component as having a power-law luminosity function only at high luminosities, $\Phi_{EVO} = \Phi_{0, EVO} L^{\alpha}$ only when $M_{K_s} \le -29.5$.
We fit the observed LoBAL fractions from the $g$ to $K_s$ bands using Model II, and obtain considerably better fits.
We show the Model II fits in Figs.~\ref{fig:thr} (right) and \ref{fig:fiv} (left), and list the fitting results in Table~\ref{tab:ifrac}.
The intrinsic fractions from the geometric component are 3.3$\pm$0.4, 5.4$\pm$0.5, and 1.3$\pm$0.3 per cent for BI-LoBALs, AI-LoBALs, and FeLoBALs, respectively,
and they are all smaller than the intrinsic fractions obtained from a pure geometric model fit (Model I).
The fractions for the evolutionary component are functions of luminosities, and the total intrinsic fractions are the sum of the two components.
The parameters of the evolutionary component, $\alpha$ and $\Phi_{0, EVO}$, are unconstrained from our data.  We list one set of possible solutions in Table~\ref{tab:ifrac} as an example.
After obtaining the best fit models, we plot the model intrinsic fractions of LoBALs as a function of $K_s$ band luminosity in Figs.~\ref{fig:fou} and \ref{fig:fiv} (right), for AI-LoBALs, BI-LoBALs (G09), and FeLoBALs, respectively.
In all three cases, we find that the models fit the observed $K_s$ band fractions well.  
The models (solid lines), which are the intrinsic fractions, represent an envelope for observed fractions. The observed fractions are below the models since they are biased with different degrees.
The $K_s$ band fractions should be closest to the model, since there is less absorption in the $K_s$ band.
The observed $K_s$ band fractions for LoBALs fainter than $K_s > -29$ are slightly below the models, which is possibly associated with the incomplete $K_s$ band data.

\subsection{Radio Properties of LoBALs}

\begin{figure*}
    \includegraphics[width=8.5truecm]{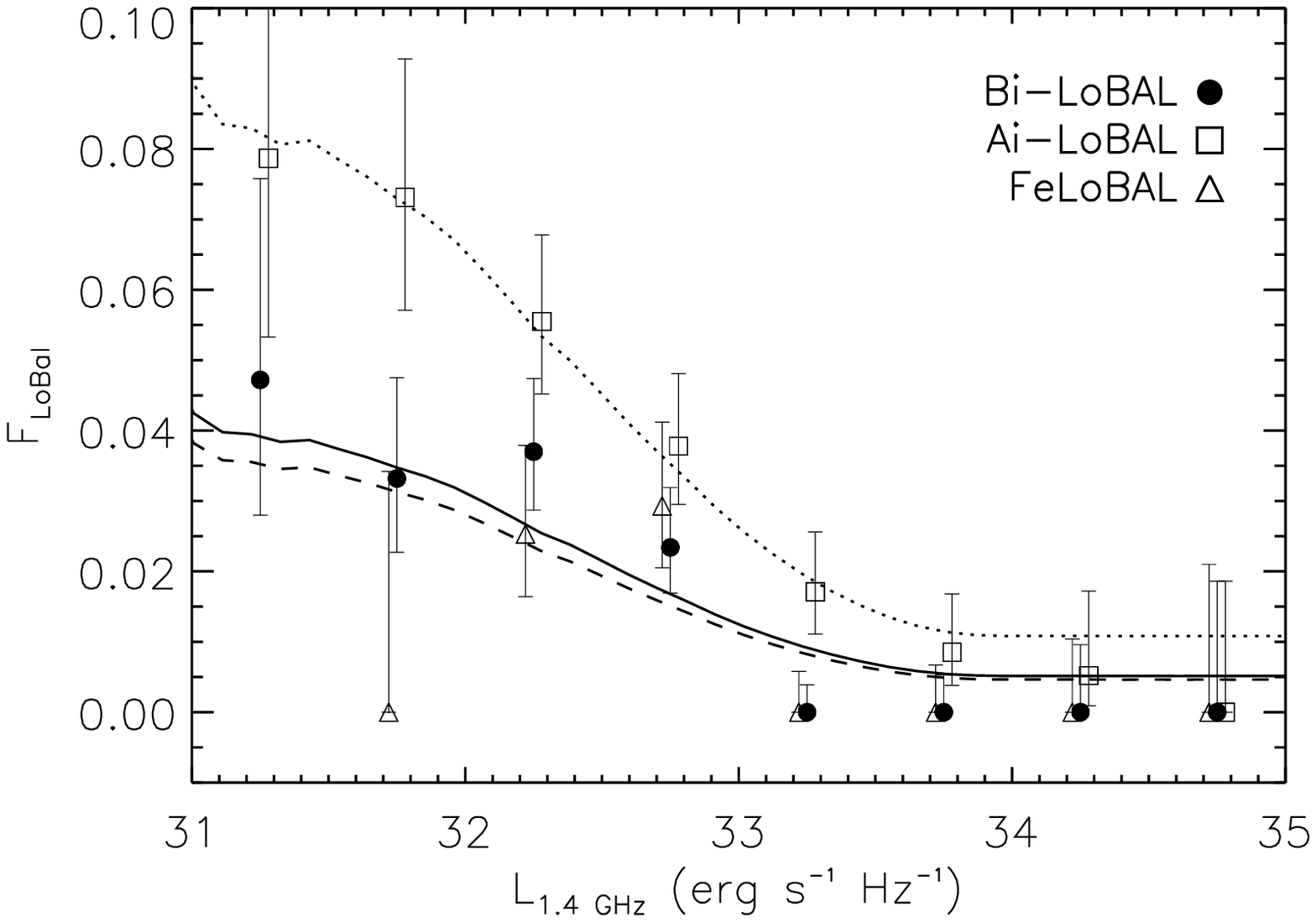}
    \includegraphics[width=8.5truecm]{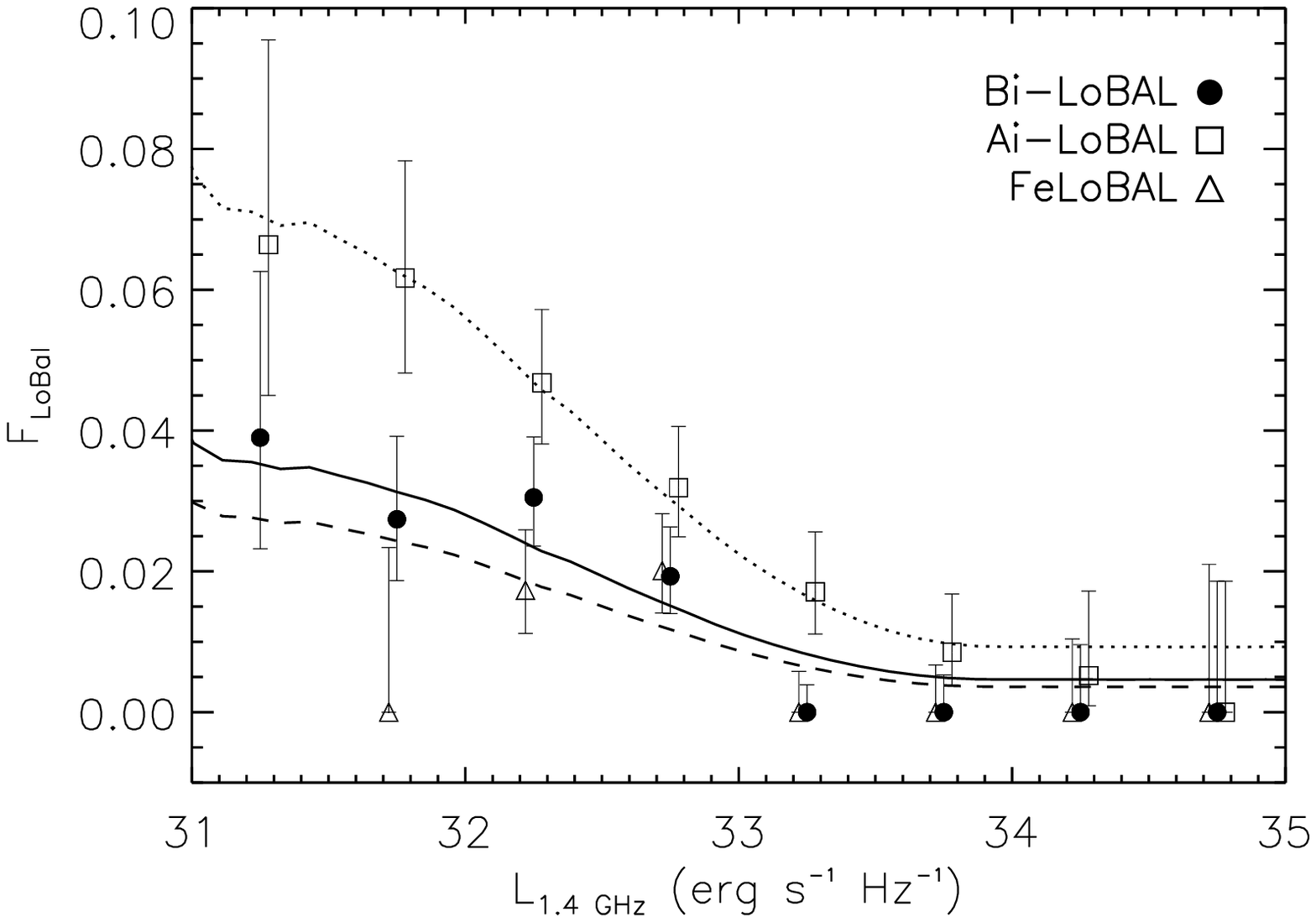}
    \caption{(left) Fractions of LoBALs and FeLoBALs as a function of the 1.4~GHz radio luminosity.  The data are slightly shifted horizontally for clarity.  At the low luminosity end, the fractions are consistent with the intrinsic fractions modelled through the NIR and optical bands.  The fractions are significantly smaller at high radio luminosities. Using a geometric model for LoBALs (solid, dotted, and dashed lines for BI-LoBALs, AI-LoBALs, and FeLoBALs, respectively), we are able to fit the trend and obtain intrinsic fractions of $4.0\pm0.7$, $8.4\pm1.0$, and $3.6\pm1.0$ per cent for BI-LoBALs, AI-LoBALs, and FeLoBALs, respectively. (right) We subtract the evolutionary component at $L_{1.4 GHz} \le 10^{33}\lumind$, and fit the modified data using a geometric model.  We find intrinsic fractions of 3.6$\pm$0.6, 7.2$\pm$0.9, and $2.8\pm0.8$ per cent for the geometric component for  BI-LoBALs, AI-LoBALs, and FeLoBALs, respectively.\label{fig:six}}
\end{figure*}

\begin{table*}
\caption{Modeling Results for the Intrinsic Fractions of LoBALs Using the FIRST Data\label{tab:ffrac}}
\begin{center}
\begin{tabular}{lccccc}
\hline
\hline
{} & Model & AI-LoBAL  & BI-LoBAL  & FeLoBAL  \\
\hline
Intrinsic Fraction (per cent) & Model I & 8.4$\pm$1.0 & 4.0$\pm$0.7 & 3.6$\pm$1.0 \\
$\chi^2/dof$ &  & 0.9/7 & 9.9/7 & 6.3/6 \\
Intrinsic Geometric Fraction (per cent) & Model II & 7.2$\pm$0.9 & 3.6$\pm$0.6 & 2.8$\pm$0.8 \\
$\chi^2/dof$ &  & 0.5/7 & 7.5/7 & 4.7/7 \\
\hline
\end{tabular}
\end{center}
\end{table*}

\begin{table*}
\caption{Final Intrinsic Fractions of LoBALs by Combining the Two Methods \label{tab:cfrac}}
\begin{center}
\begin{tabular}{lccccc}
\hline
\hline
{} & Model & AI-LoBAL  & BI-LoBAL  & FeLoBAL  \\
\hline
Intrinsic Fraction (per cent) & Model I & 7.1$\pm$0.6 & 4.0$\pm$0.5 & 2.1$\pm$0.3 \\
Intrinsic Geometric Fraction (per cent) & Model II & 5.8$\pm$0.4 & 3.4$\pm$0.3 & 1.5$\pm$0.3 \\
\hline
\end{tabular}
\end{center}
\end{table*}

We show the fraction of LoBALs as a function of the 1.4~GHz radio luminosity in Fig.~\ref{fig:six} (left).
In the $0.5 \le z \le 2.15$ redshift range, we find 2650, 95, and 49 matches in FIRST for the SDSS quasars, AI-LoBALs, and BI-LoBALs, respectively.
In the $1.19 \le z \le 2.24$ redshift range, we find 1621 and 19 matches in FIRST for the SDSS quasars and FeLoBALs, respectively.
We find that the fraction of LoBALs decreases with increasing radio luminosities.  This is true in all of the sub-samples of LoBALs (BI-LoBALs, AI-LoBALs, AINB-LoBALs, and FeLoBALs).
We do not show the AINB-LoBAL fractions in Fig.~\ref{fig:fou} for clarity, but the numbers can be obtained by subtracting the BI-LoBAL fractions from the AI-LoBAL fractions.
In addition, the LoBAL fraction at the low radio luminosity end is consistent with the intrinsic fraction obtained in Section~\ref{sec:intr}.
This is not unexpected since there is also little obscuration in the radio band.
These two features are similar to BALQSOs (Shankar et al.\ 2008a).

The decrease in the LoBAL fractions as a function of radio luminosity can be easily explained under a geometric model for LoBALs, if we assume that LoBALs are viewed in the lines of sight close to the accretion disc, while the boosted radio emission is viewed close to the polar direction.
We test this model following Shankar et al.\ (2008a).  We assume that the radio emission is composed of one weaker uniform component and another strong, beamed component at the polar direction.
LoBALs are modelled to be within a solid angle close to the equatorial plane, where the open angle is a free parameter.
The rest of the parameters, such as the slope of the luminosity function, Lorentz factor, maximum beaming angle, are all fixed according to the literature (Urry \& Padovani 1991; Urry \& Padovani 1995; De Zotti et al.\ 2005; Richards et al.\ 2006; Jiang et al.\ 2007; Padovani et al.\ 2007).
Please refer to Shankar et al.\ (2008a) for the detailed model.
Using this model, we find good fits to the data with $\chi^2/dof$ = 1.4, 0.12, and 1.0 for BI-LoBALs, AI-LoBALs, and FeLoBALs, respectively, and the corresponding intrinsic fractions of these objects are $4.0\pm0.7$, $8.4\pm1.0$, and $3.6\pm1.0$ per cent in quasars.
These fractions are consistent with the LoBAL fractions obtained using a pure geometric model (Model I) in Section~3.2; however, they are higher than those obtained using a two components model (Model II).

Since the infrared and optical data favour an additional evolutionary component for LoBALs, we try to incorporate this component in our radio model.
From the radio flux versus the $K_s$ magnitude plots in Figs.~\ref{fig:one} and \ref{fig:two}, it is likely that the $K_s$ bright LoBALs have relatively faint radio fluxes.
Since there is little absorption in the radio band and our model shows that the measured fractions at low radio luminosities are close to the intrinsic fractions, we can subtract the evolutionary component assuming the ratio of the evolutionary component to the geometric component is a constant at low radio luminosities.
Based on this, we subtract the evolutionary component in the LoBAL fractions at low radio luminosities ($L_{1.4 GHz} \le 10^{33}\lumind$) using the ratio between the model intrinsic fraction for the geometric component in Model II and the model intrinsic fraction in Model I in section Section~3.2.
The resulting data are shown in Fig.~\ref{fig:six} (right), where we find lower fractions compared to the left panel at $L_{1.4 GHz} \le 10^{33}\lumind$.
We fit these data using a pure geometric model as above, and find intrinsic fractions of 3.6$\pm$0.6, 7.2$\pm$0.9, and $2.8\pm0.8$ per cent for the geometric components for BI-LoBALs, AI-LoBALs, and FeLoBALs, respectively.  We also sightly improve the fitting statistics.  The fitting results are listed in Table~\ref{tab:ffrac}.

\section{Summary and Discussion}
\label{sec:discuss}
We find significantly high fractions of LoBALs in the quasar population compared to the values obtained using optical data only.
For example, the BI-LoBAL and AI-LoBAL fractions in the optical data were measured as 0.55 and 1.31 per cent (T06), while our results are 5--7 times larger.
Although the final intrinsic fractions depend on the choice of catalogues, the overall trend is found to be the same.
For example, we perform a similar analysis to the BI-LoBALs from the G09 sample, and also find large intrinsic LoBAL fractions.
Although there is a systematic offset between the BI-LoBAL fractions from the T06 and G09 samples, the overall trend for the observed fractions is the same.
Our intrinsic fractions are obtained using two independent methods, one from the NIR and optical data and the other from the radio data, and the results are mutually consistent between the two methods.
Combining the estimates from the two methods using the least square (minimum variance) method, we find that the intrinsic fractions for BI-LoBALs, AI-LoBALs, and FeLoBALs are $4.0\pm0.5$, $7.1\pm0.6$, and $2.1\pm0.3$ per cent, respectively, using a pure geometric model.  For our hybrid model with both the geometric and evolutionary components, which fit the data better, the corresponding intrinsic fractions for the geometric component are $3.4\pm0.3$, $5.8\pm0.4$, and $1.5\pm0.3$ per cent.
The intrinsic fractions for the evolutionary component for LoBALs are functions of luminosities, and the total intrinsic fractions of LoBALs are the sums of the two components.
The final combined estimates of the intrinsic LoBAL fractions are listed in Table~\ref{tab:cfrac}.
The results are not unexpected considering the large obscuration observed in LoBALs (e.g., Sprayberry \& Foltz 1992).
Although we find significantly larger intrinsic fractions of LoBALs, they still represent a small portion of the total population.
Compared with the intrinsic fractions of $20\pm2$ and $43\pm2$ per cent for BI-BALQSOs and AI-BALQSOs (Dai et al.\ 2008a), the LoBALs are about 20 per cent of BALQSOs.

Our method of calculating the intrinsic fractions of BALQSOs still depends on the completeness of optical quasar surveys.  The fraction of quasars that do not enter the optical surveys was not be accounted in this or our previous papers.  It was estimated that SDSS is about 90 per cent complete at $z < 2.2$ and $i < 19.1$~mag (Richards et al.\ 2002).  The remaining 10 per cent of quasars, which are thought to be highly obscured, could potentially all be LoBALs or FeLoBALs.  Thus, in most optimistic estimates, the fractions of LoBALs or FeLoBALs can reach $\sim$15 per cent for the geometric component.  However, the nature of the obscured quasars are still uncertain, and we are not sure whether the ultra-violet spectra of these quasars still show broad absorption lines, if the continua of these quasars are mostly obscured.
Therefore, the fractions quoted in our paper represent conservative estimates based on observations.

The LoBAL fractions in the radio band are particularly interesting, since we find that the LoBAL fractions decrease with increasing radio luminosities.  This confirms the early result of Becker et al.\ (2000) with about a dozen LoBALs.
The trend is similar to that found in the total BALQSO population (Shankar et al.\ 2008a).
This trend found in both the total BALQSO population and LoBALs suggests that
the majority of LoBALs and BALQSOs can be united under a similar physical scheme.
In Shankar et al.\ (2008a), we favoured a geometric model to interpret the trend.
Applying the geometric model of Shankar et al.\ (2008a) to LoBALs, we can successfully reproduce the data for BI-LoBALs, AI-LoBALs, and FeLoBALs.
However, it is problematic under a pure evolutionary model to explain the radio-luminosity/LoBAL fraction trend as argued by Shankar et al.\ (2008a).  The main question is why quasars spend the same fraction of time as BALQSOs in the radio-loud stage and radio-quiet stage.
Therefore, we argue that the majority of BALQSOs and LoBALs can be understood in a geometric model.
Polar BALQSOs/outflows (Zhou et al.\ 2006; Ghosh \& Punsly 2007; Dai et al.\ 2008b) present a challenge to our results; however, these objects are rare and we are uncertain about their implications to the total quasar population.
A modification of the geometric model to have both disc and polar outflows (e.g., Borguet \& Hutsem{\'e}kers 2010) may be needed to incorporate these objects.
We also note that LoBALs dominate this population of polar BALQSO candidates (Ghosh \& Punsly 2007).

There are other indications, such as the association with ULIRGs, radio spectra, and large covering fractions from spectral modeling, arguing that LoBALs belong to an earlier evolutionary stage of quasar population.
However, most of these studies are based on a small sample size and may not extrapolate to the whole LoBAL population.
Urrutia et al.\ (2009) studied the fraction of LoBALs in the dust reddened quasars at high redshift, finding that all except one are LoBALs, supporting the young nature of LoBALs.
However, the authors also noted that their selection method may be biased favouring LoBALs since they are associated with large dust reddening.
In our study, we find LoBALs and BALQSOs are similar in most aspects, except that the fraction of LoBALs increases with increasing NIR luminosities.
This is not consistent with BALQSOs in general, because the BALQSO fractions are mostly constant with increasing NIR luminosities for AI-BALQSOs (Dai et al.\ 2008a).
At the NIR luminous end, the observed LoBAL fraction is higher, although with large uncertainties, than the intrinsic fraction that we obtain for a pure geometric model.

It is possible that a portion of NIR luminous LoBALs are special compared to the rest of the population, and at an early evolutionary stage of the quasar cycle.
This will reconcile some observations supporting the young quasar interpretation for LoBALs.
We add an evolutionary LoBAL component to our model, and generally obtain a better fit to the data.
However, the parameters of this evolutionary component are unconstrained from the data.
The combination of a short early evolution model, with a large covering fraction close to 100 per cent, and a subsequent, longer geometric model, with covering fractions consistent with the intrinsic fraction measured in this paper and Dai et al.\ (2008a), could simultaneously account for many of the differences and similarities between LoBALs and HiBALs.
The longer-lived geometric model mainly sets the intrinsic fractions of BALQSOs of various species and the measured fraction as a function of their radio luminosities.
If there is some small spherical outflow component at early times, this might also explain the predominance of
LoBALs among the rare polar BALQSOs.
Deeper IR and radio surveys are needed to increase the sample size and confirm this claim.

\section*{Acknowledgments}
We acknowledge the anonymous referee for helpful comments.
F.\ S.\ acknowledges support from the Alexander von Humoboldt Foundation.

\label{lastpage}
\end{document}